\documentclass[aps,prl,reprint,preprintnumbers,showpacs,amsmath,amssymb,letterpaper]{revtex4-1}

\usepackage{graphicx}
\usepackage{dcolumn}
\usepackage{bm}
\usepackage{color}

\begin{document}

\preprint{This article has been submitted to Physical Review Letters. When published, it will be available at \textcolor{blue}{ http://prl.aps.org/}.}


\title{The nonlinear dispersion relation of geodesic acoustic modes}

\author{Robert Hager}
 \email{robert.hager@ipp.mpg.de}
\author{Klaus Hallatschek}%
 \email{klaus.hallatschek@ipp.mpg.de}
\affiliation{%
Max-Planck-Institut f\"ur Plasmaphysik\\
Boltzmannstra{\ss}e 2, D-85748, Garching, Germany
}%

\date{\today}


\begin{abstract}
The energy input and frequency shift of geodesic acoustic modes (GAMs) due to turbulence in tokamak edge plasmas are investigated in numerical two-fluid turbulence studies. Surprisingly, the turbulent GAM dispersion relation is qualitatively equivalent to the linear GAM dispersion but can have drastically enhanced group velocities.
In up-down asymmetric geometry the energy input due to turbulent transport may favor the excitation of GAMs with one particular sign of the radial phase velocity relative to the magnetic drifts and may lead to pulsed GAM activity.
\end{abstract}


\maketitle

%
\paragraph{Introduction.---}Geodesic acoustic modes (GAMs), oscillating plasma flows associated with radial gradients of the electrostatic potential \cite{winsor_gamfreq}, are an ubiquitous phenomenon in tokamak edge plasmas \cite{conway_iphase2}. By shearing the turbulence, GAMs are able to modulate the turbulent transport and to reduce the saturation intensity of the turbulence \cite{hall_transport}.

Using numerical two-fluid turbulence studies with the code \textsc{nlet} \cite{hall_nlet}, we investigate the energy input and frequency shift of GAMs due to their interaction with the turbulence.
From the radial structure of global GAM eigenmodes \cite{itoh_eigenmode} in non-local (non-Boussinesq) computations we deduce a \emph{nonlinear}, i.e. turbulence induced, dispersion relation. Although the observed dispersion agrees qualitatively with the ones obtained from linearized two-fluid equations \cite{hager_gam,hager_gam_asym}, the corresponding group velocities tend to be surprisingly high.
For realistic edge parameters, the global eigenmodes have a radial extent of several centimeters and could account for the frequency plateaus in ASDEX Upgrade \cite{conway_gamfreq}, which deviate from the expected $T^{1/2}$ dependence of acoustic frequencies.
In turbulence runs with broken up-down symmetry we observe periodic bursts of turbulence and GAM activity.
They can be traced back to an energy transfer from the turbulence to the flows which is odd in the radial wavenumber and absent with symmetry.
This behavior resembles strikingly the experimentally observed pulsation during the I-phase in ASDEX Upgrade \cite{conway_iphase2} or the periodic turbulence suppression in NSTX \cite{zweben_quiet_periods}.

%
\paragraph{Computational Setup.---}
We shortly summarize the basic computational setup of the turbulence runs.
The ratio of the turbulence- to the background-gradient scale lengths is $\lambda^{-1} \equiv L_\perp / L_n \propto \rho^\star$, where $L_\perp$ is a characteristic turbulence scale length perpendicular to the magnetic field, $L_n^{-1} \equiv \partial_r \ln n$, and the flux label $r$ is the minor radius at the outboard midplane. (A ratio of $\lambda \sim 1-50$ is typical for the edge region.)
The computational domain covers $490$ sound gyro radii $\rho_{s}\equiv (\gamma m_i (T_e+T_i))^{1/2}/(e B)$ radially and poloidally.
$m_i$ is the ion mass, $T_{e,i}$ the electron and ion temperatures, $e$ the elementary charge, and $B$ the magnetic field. The adiabatic exponent $\gamma$ is $5/3$.
Furthermore, we choose $\eta_i\equiv L_n/L_{T_i} = 2.4$, $\eta_e\equiv L_n/L_{T_e} = 0$, $\epsilon_n\equiv 2L_n/R_0=0.05$, $\alpha_d=0.5$, $\epsilon_v=0$, and $\gamma_p=0$, where $\alpha_d$ is the ratio of the drift frequency to the ballooning growth rate, $\gamma_p$ represents magnetic pumping, and $\epsilon_v$ the sound speed (for definitions see \cite{hall_nlet}). $R_0$ is the major radius at the outboard midplane.
The electrons are isothermal and treated non-adiabatically.
Density and temperature are normalized to their values in the middle of the radial computational domain, i.e. $(\tilde{n},\tilde{T}_{e/i}) \equiv \lambda (n,T_{e/i})/(n_{0},T_{e/i,0})$, and $\tau \equiv T_{i,0}/T_{e,0} =1$. The background density and temperature profiles are linear, $\tilde{n}_b=\lambda - 0.25 r$ and $\tilde{T_i}_b=\lambda - 0.62 r$. Thus, the local GAM frequency at zero radial wave number $\omega_{GAM,0}(r)$ varies by roughly $\pm 0.3 \omega_{GAM,0}(r=0)$, where $r=0$ marks the middle of the radial domain. The magnetic field is normalized to its value at the outboard midplane. With the time units $t_0=R_{eff}/(\sqrt{2}c_s)$, where $c_s \equiv (\gamma (T_e+T_i)/m_i)^{1/2}$ is the sound speed, $1/R_{eff} \equiv (2 \langle C(\theta)^2 \rangle)^{1/2}/R_0$, $C(\theta) \equiv (R_0 |\nabla r|/B) (\mathbf{b} \times \nabla \ln B) \cdot \hat{\mathbf{r}}$ and $\langle \dots \rangle \equiv \oint \dots B^{-1} \mathrm{d}l_\parallel / \oint B^{-1} \mathrm{d}l_\parallel $ is the flux-surface average, $\omega_{GAM,0}(r=0)=1$.

%
\paragraph{Nonlinear GAM dispersion.---}
\begin{figure}
 \centering
 \includegraphics[width=3.0in]{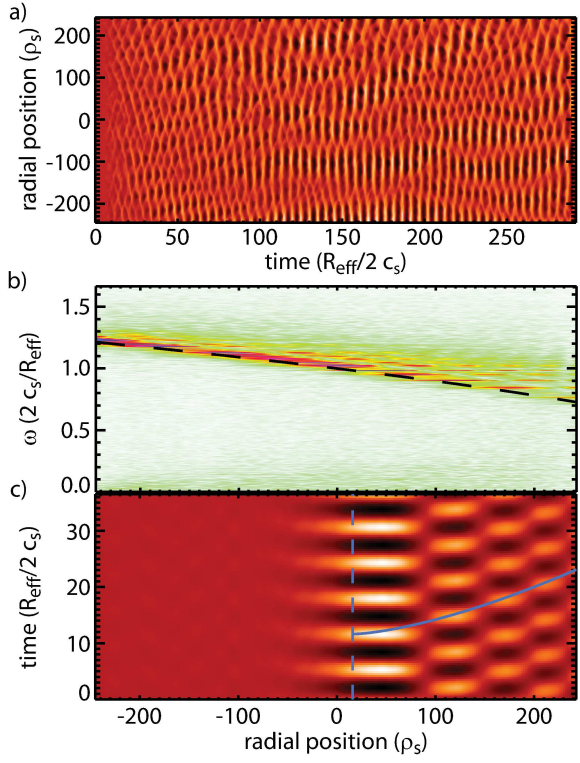}
 \caption{\label{fig:flowdata_bob65nl_ha} (color online) Results of a non-local NLET turbulence study with slow parameter variation, $\lambda \approx 200$. a) Flux-surface averaged poloidal $\mathbf{E} \times \mathbf{B}$ velocity. b) Temporal Fourier transform (linear color scale) of a) with the local GAM frequency indicated by the dashed line. c) Band-pass filtered flow profile at $\omega=0.98$ with a fit to a curve of constant phase according to Eq. \eqref{eqn:phaseline}.
 }
\end{figure}
Figure \ref{fig:flowdata_bob65nl_ha} shows the result of a nonlocal ($\lambda \approx 200$) ITG turbulence computation with circular flux-surfaces, which displays pronounced GAM activity. The radial variation of the GAM frequency caused by non-locality is obvious in the corresponding $\mathbf{E}\times\mathbf{B}$-flow profile [Fig. \ref{fig:flowdata_bob65nl_ha} a)].
Its temporal Fourier transform [Fig. \ref{fig:flowdata_bob65nl_ha} b)] reveals that GAMs with frequency $\omega_{GAM,0}(r_0)$ (indicated by the dashed line) radiate significantly outwards to $r>r_0$ but disappear for $r<r_0$.
The deviations from the local GAM frequency reach up to $30$ \%.
The application of a narrow band pass filter [Fig. \ref{fig:flowdata_bob65nl_ha} c)] shows that the flow profile consists of \emph{global} GAM eigenmodes.
Such an eigenmode can be described by a WKB wave packet with a local wave number $k_r$ obeying $\omega_{GAM}(r, k_r) = const$.
To relate the observed deviations from $\omega_{GAM,0}(r_0)$ to the nonlinear GAM properties, we choose the ansatz $\omega_{GAM}(r, k_r) = \omega_{GAM,0}(r) (1+ \alpha k_r^2)$ for the dispersion relation.
The resulting solution for $k_r$ is
\begin{equation}\label{eqn:wavenumber}
 k_r (r)  \approx \sqrt{ \frac{1}{\alpha L_\omega (r_0)} \left( r_0-r \right) },
\end{equation}
where $L_\omega$ is the gradient length of $\omega_{GAM,0}$. Depending on the sign of $\alpha$, $k_r$ is imaginary on one side of the flux-surface $r=r_0$ and real on the other. $k_r(r_0)=0$ implies the existence of a reflection layer at $r_0$.
Using Eq. (\ref{eqn:wavenumber}) and the condition $\dot{r} = \omega_{GAM,0}(r_0)/k_r$ for the phase velocity yields the curve of constant phase
\begin{equation}\label{eqn:phaseline}
 t(r) = \frac{2}{3} \sqrt{-\frac{1}{\alpha L_\omega(r_0) \omega_{GAM,0}^2(r_0)}} \left( r-r_0 \right)^{3/2} + t_0.
\end{equation}
A least squares fit for the band-pass filtered flow profile in Fig. \ref{fig:flowdata_bob65nl_ha} c) with $\omega_{GAM,0}(r_0)=0.98$ and the corresponding reference radius $r_0$ indicated by the dashed line yields $\alpha \approx 41 \rho_s^2$, roughly a hundred times the value observed in the absence of turbulence ($|\alpha| \sim \rho_s^2/2$ \cite{hager_gam}).
A comparison with the corresponding local run in Fig. \ref{fig:nonlin_dispersion_results} a) confirms the obtained dispersion relation.
The above values for $\alpha$ can be used to estimate the radial extent of a global eigenmode whose radial structure is on one hand given by an Airy function \cite{itoh_eigenmode}. On the other hand the turbulence excites GAMs only in a limited range of radial wave numbers, whose distribution is assumed to be a Gaussian of width $\sigma_k$ centered at $k_r=k_0$.
Since the Airy function $\mathrm{Ai}(x)$ drops only as $x^{1/4}$, the wave number distribution represents the dominant limitation of the mode width. The estimate resulting from Eq. (\ref{eqn:wavenumber}) is therefore $\delta r \lesssim 4 \alpha L_{\omega} \sigma_k k_0$.
On a basis of many turbulence computations we choose $k_0 \rho_s \sim 0.1$ and $\sigma_k \rho_s \sim 0.05$.
In tokamak edge plasmas $L_\omega \sim L_T \sim 0.01\dots 0.1 R$. Thus, assuming a major radius of $R \sim 1\, \mathrm{m}$, we obtain mode widths of $\delta r \lesssim 0.1 \dots 1 \,\mathrm{mm}$, i.e. of the order of the gyro radius for $\alpha = \rho_s^2/2$. In contrast, for $\alpha = 41 \rho_s^2$ $\delta r \lesssim 0.8\dots8.2 \,\mathrm{cm}$. The corresponding group velocities are of the order of the ion magnetic drifts and the diamagnetic drift, respectively. Therefore, the width of nonlinear eigenmodes can be comparable to the scale length of the frequency plateaus reported in \cite{conway_gamfreq}.

Now, we develop an understanding of the turbulence induced frequency shift.
The turbulent source terms are particularly transparent if we write the time evolution of the GAM in terms of a state vector $\Psi \equiv (p_{GAM},v_{GAM})$ as
\begin{align}\label{eqn:gam_eqs}
 \partial_t \Psi = \delta\Psi_{lin}+\delta\Psi_{nl},
\end{align}
which can be derived from the two-fluid system in Ref. \cite{hall_nlet}. Here, $\delta\Psi_{lin} \equiv (v_{GAM},-p_{GAM})$ describes the linear time evolution of the GAM and $\delta\Psi_{nl}\equiv (s_\Gamma+s_{dia},s_\Pi)$ the nonlinear source terms. The GAM related part of the pressure fluctuations $p$ is $p_{GAM} \equiv \langle \tilde{\mathrm{C}}(\theta) p \rangle$ with $\tilde{\mathrm{C}} \equiv \mathrm{C}/\langle \mathrm{C}^2 \rangle^{1/2}$. The poloidal velocity component of the GAM is defined as $v_{GAM} \equiv \langle \partial_r \phi\rangle = \langle v_{E,\theta}/\xi \rangle$, where $v_{E,\theta}$ is the poloidal component of the $\mathbf{E}\times\mathbf{B}$ velocity $\boldsymbol{v}_{E}= \xi \mathbf{b} \times \nabla_\perp \phi$. Here, $\phi$ is the potential fluctuation and $\xi \equiv (1+\tau)/(2 (1+\gamma \tau))$. Pressure and electric potential perturbations, length, and time are normalized such that the GAM frequency $\omega_{GAM,0}=1$ and the free energy \cite{hager_gam} is $E=p^2+(\partial_r \phi)^2$.
The divergence of the Reynolds stress is given by $\mathrm{s_\Pi} \equiv  - \langle \partial_r (v_{E,\theta} v_{r,ion}) \rangle$, the divergence of the up-down asymmetric component of the turbulent transport by $\mathrm{s_\Gamma} \equiv - \langle \partial_r  (\tilde{\mathrm{C}}(\theta) v_{E,r} p) \rangle$, and the background diamagnetic velocity by $s_{dia} \equiv \langle \alpha_\tau \partial_r (p_i + \beta_\tau T_i) \rangle$ with $\alpha_\tau \equiv \tau/(1+\tau)$ and $\beta_\tau=\gamma\tau (1+\gamma \tau)$ and the ion pressure $p_i \equiv n+T_i$.
The energy stored in the GAM corresponds to the squared length of the state vector $\Psi$. Obviously, $\delta\Psi_{lin}$ is orthogonal to $\Psi$ and is responsible for the GAM oscillation. Components of $\delta\Psi_{nl}$ changing the length of $\Psi$ alter the GAM energy, the components orthogonal to $\Psi$ cause frequency shifts.
Using the basis vectors $\hat{e}_E \equiv (1/|\Psi|) \Psi$ and $\hat{e}_\varphi \equiv (1/|\Psi|) (v_{GAM},-p_{GAM})$ one obtains the nonlinear energy input $\partial_t E_{nl} = (\delta\Psi_{nl} \cdot \hat{e}_E)^2$ and the frequency shift $\delta\omega = \langle (1/|\Psi|) \delta\Psi_{nl} \cdot \hat{e}_\varphi \rangle_t$ with the time average $\langle \dots \rangle_t$.
For the Fourier mode with $\omega=0.98$ [Fig. \ref{fig:flowdata_bob65nl_ha} c)], this splitting of the nonlinear terms into energy inputs and frequency shifts clearly shows that the observed frequency shift is due to the coupling to the up-down asymmetric component of the turbulent transport $s_\Gamma$ alone, while the Reynolds stress and the diamagnetic velocity do not contribute [Fig. \ref{fig:nonlin_dispersion_results} b)].
The increase of the GAM frequency (and the phase velocity) due to the asymmetric transport $\Gamma \equiv v_{E,r} p$ requires $\Gamma$ to be in phase with $p_{GAM}$.
A reduction of the resistivity results in reduced GAM group velocities, indicating a strong dependence of this effect on the ballooning character of the turbulence.

\begin{figure}[t]
 \centering
 \includegraphics[width=3.0in]{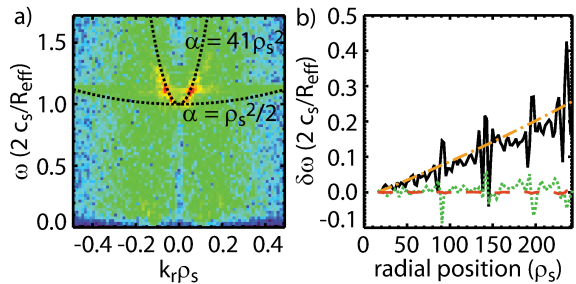}
 \caption{\label{fig:nonlin_dispersion_results} (color online)a) Spectrum of the flux-surface averaged poloidal $\mathbf{E} \times \mathbf{B}$ velocity (logarithmic color scale) of a local NLET turbulence study otherwise equivalent to the one shown in Fig. \ref{fig:flowdata_bob65nl_ha} (using the parameters at $r=0$) and dispersion relations $\omega_{GAM}(r,k_r)=\omega_{GAM,0}(r) (1+\alpha k_r^2)$ with $\alpha=\rho_s^2$ (linear dispersion) and $\alpha=41 \rho_s^2$ b) Nonlinear frequency shifts due to the couplings between turbulence and GAMs for the Fourier mode $\omega=0.98$ of Fig. \ref{fig:flowdata_bob65nl_ha} a): frequency shift necessary for the radial mode structure in Fig. \ref{fig:flowdata_bob65nl_ha} c) (dashed-dotted), shift due to the up-down asymmetric component of the turbulent transport (solid), Reynolds stress by the (dotted), and diamagnetic velocity (dashed). 
}
\end{figure}
%

\begin{figure}
 \centering
 \includegraphics[width=3.0in]{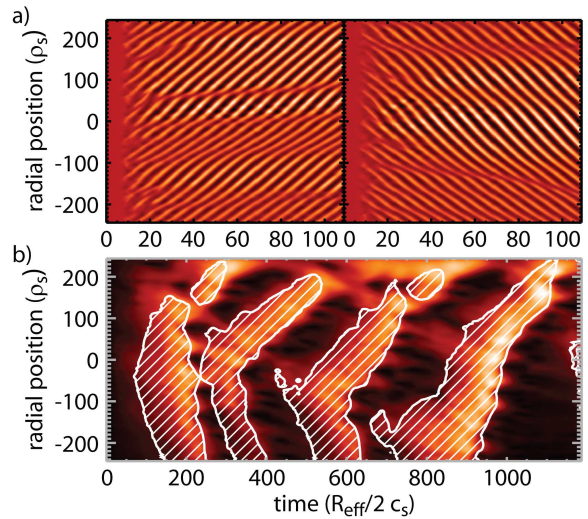}
 \caption{\label{fig:xpoint_influence} (color online) Flux-surface averaged poloidal $\mathbf{E} \times \mathbf{B}$ flows of NLET turbulence studies using the same parameters as in Fig. \ref{fig:flowdata_bob65nl_ha} but with a single-null geometry. a) Local run. Left: Ion magnetic inhomogeneity drift $\mathbf{v}_d$ directed away from the X-point. Right: $\mathbf{v}_d$ directed towards the X-point. b) Non-local run with $\lambda \approx 200$ and $\mathbf{v}_d$ towards the X-point. Color-coded: short-time-RMS of $\langle v_{e,\theta} \rangle$. Contours: turbulence intensity. Within the shaded areas, the turbulence intensity is above $70$ \% of its RMS average.
 }
\end{figure}

%
\paragraph{Pulsed GAM activity.---}
We investigate the effect of up-down asymmetry of the magnetic geometry in local edge-turbulence computations using single-null geometry with the same parameters as in the previous section but with adiabatic electrons.
The flux surface is constructed with the magnetic field of five toroidal current loops representing the plasma current and the currents in the shaping coils \cite{hager_gam_asym}. The poloidal flux is $\Psi = \sum_i A_i = 0.999 \Psi_{sep}$, where $A_i=a_i/2 \ln ((R_a+r_i)^2+(Z+z_i)^2)$ and $\Psi_{sep}$ is the poloidal flux at the separatrix.
The coefficients $a_i$, $z_i$, and $r_i$ measure the current of conductor $i$, its position on the (vertical) $Z$-axis, and its radial position.
The radius of the magnetic axis is $R_a$. For consistency, the coefficients are chosen such that the forces on the plasma current exerted by the four shaping coils add to zero, i.e. $\sum_i a_i /(r_i^2+z_i^2)^{1/2}=0$. We choose $r_0=z_0=0$, $a_0=1$, $r_1=0$, $z_1=m$, $a_1=-m$, $r_2=-r_3=z_2=z_3=m/\sqrt{2}$, $a_2=a_3=\sqrt{2} m$, $r_4=0$, $z_4=-1$, and $a_4=1$ with $m=2$ for lower ($z_i \rightarrow -z_i$ for upper) X-point configuration. Ion curvature and $\nabla B$ drifts are directed in positive $Z$-direction. In local turbulence runs, the system saturates in both configurations, upper and lower X-point, by exciting GAMs.
However, a striking difference is the asymmetry of the nonlinear GAM drive with respect to the sign of the phase velocity of the GAM, which is obvious in the $\mathbf{E} \times \mathbf{B}$-flow profiles in Fig. \ref{fig:xpoint_influence} a). Only GAMs with radially outward phase velocity are excited with the curvature drift directed away from the X-point whereas the phase velocity is radially inward with the curvature drift directed towards the X-point.
The characteristic radial scale length of the GAMs is the same in both cases. The corresponding $(k_r,\omega)$-spectra reveal that the GAM activity is concentrated close to the linear dispersion. Since the linear group velocity of the GAM in single-null configuration is in direction of the ion curvature drift at a location opposite to the X-point \cite{hager_gam_asym}, one can say that GAMs are excited such that the phase velocity is in direction of the group velocity.
Including the parameter variation ($\lambda \approx 200$) the GAM flow amplitude and the turbulence intensity become correlated and pulsed at a frequency much smaller than the GAM frequency [Fig. \ref{fig:xpoint_influence} b)].
Detailed analysis of the poloidal flow shows that the characteristic radial scale $k_{r,i}$ of the GAM in the initial phase ($t < 100$) is the same as in the local studies. The phase velocity is negative. However, due to the frequency gradient the wave fronts of the flows are being tilted in the $r-t$-plane and $k_r$ decreases. In response, the GAM-turbulence equilibrium readjusts itself towards transport levels even higher than in computations with artificially suppressed zonal flows.
At $k_r=0$ the GAM and turbulence intensity reach a maximum and then drop for $k_r>0$ as the GAM is damped.
After this burst, the described cycle repeats. The pulse frequency obviously depends on the GAM frequency gradient and on the initially growing wave number, i.e. $\omega_b \approx 2 \pi/(k_{r,i} \rho_s) \partial_r \omega_{GAM,0}(r)$.
A range of $\omega_b \approx 0.01 \dots 0.5 \omega_{GAM}$ seems to be readily realizable by adjusting $\partial_r \omega_{GAM,0}(r)$ and assuming $k_{r,i} \rho_s \sim 0.1$.
For opposite sign of the ion curvature drift, the GAM wave number is restricted to $k_r > k_{r,i}$. In this case, the response of the turbulence intensity on the GAM wave number turns out to be weak, and the pulsing is subdominant.

On the basis of Eq. (\ref{eqn:gam_eqs}), we investigate the dependence of the GAM-turbulence interaction on the sign of the phase velocity. 
Since the Reynolds stress turned out not to be important for the present parameters, we drop $s_\Pi$. Moreover, we approximate $s_{dia} \approx \alpha_\tau \partial_r \langle p \rangle$ because the temperature term in the diamagnetic source term does not change the result qualitatively.
The source term due to the turbulent transport $\Gamma$ can be written as $s_\Gamma= -\partial_r \langle C \Gamma\rangle$.
The transport itself is expressed as $\Gamma = \Gamma_1 + \Gamma_2$, where $\Gamma_1 \approx \hat{\Gamma}_1 (\theta) (1 - \tau_{dc} \partial_t) v_{GAM}$ (for the Galilei invariance of this term see Ref. \cite{hall_transport}), and $\Gamma_2 \approx \hat{\Gamma}_2 (\theta) (1 - \tau_{dc} \partial_t) \partial_r v_{GAM}$ \cite{hall_transport,itoh_hallatschek_gam_excitation}. Here, $\tau_{dc}$ represents the turbulence decorrelation time, which sets the time scale for the reaction of the turbulence to changes of the shear flow.
The poloidal structure of $\Gamma_i$ is contained in the function $\hat{\Gamma}_i (\theta)$.
Furthermore, $\partial_t \langle p \rangle \approx -\partial_r \langle \Gamma \rangle$ and $v_{GAM} \approx -i (\omega_{GAM,0}/\omega_{GAM}) p_{GAM}$, where $\omega_{GAM,0}$ is the linear, $\omega_{GAM}$ the (complex) nonlinear GAM frequency.
Note that  $\Re(\omega_{GAM}) \approx \omega_{GAM,0}=1$ and $\Re(\omega_{GAM}) \gg \Im(\omega_{GAM})\equiv \omega_i$. Inserting the expressions given above into Eq. (\ref{eqn:gam_eqs}), the growth rate of $p_{GAM}$ evaluates to
\begin{equation}\label{eqn:growth_rates}
 \omega_i \approx \frac{1}{2} \left( - k_r \left\langle \tilde{\mathrm{C}} \hat{\Gamma}_1 \right\rangle + k_r^2 \alpha_\tau \left\langle \hat{\Gamma}_1 + \tau_{dc} \tilde{\mathrm{C}} \hat{\Gamma}_2 \right\rangle \right) +\mathrm{O}(k_r^3).
\end{equation}
Only the term linear in $k_r$, which originates from $s_\Gamma$, can be responsible for the asymmetry of the phase velocity of the turbulent GAM excitation.
In up-down symmetric magnetic geometries $\tilde{\mathrm{C}}(\theta)$ is up-down antisymmetric (e.g. $\sin (\theta)/\sqrt{2}$ for circular flux-surfaces). Since empirically $\Gamma_1$ is approximately symmetric with respect to the outboard midplane of the tokamak and positive, $\langle \tilde{\mathrm{C}} \hat{\Gamma}_1 \rangle=0$ and the related growth rate is zero. However, in single-null geometry, $\tilde{\mathrm{C}}(\theta)$ becomes very small close to the X-point. Hence, $\langle \tilde{\mathrm{C}} \hat{\Gamma}_1 \rangle \neq 0$ and an asymmetry in the phase velocity can arise in Eq. (\ref{eqn:growth_rates}). With $\mathbf{b} \times \nabla\ln B$ (the direction of ion magnetic drifts) directed upwards -- as is convention in \textsc{nlet} -- and upper (lower) single-null geometry, $\hat{\mathrm{C}}(\theta)$ is positive (negative) opposite to the X-point. Thus, $\langle \tilde{\mathrm{C}} \hat{\Gamma}_1 \rangle$ becomes positive (negative), and the sign of the asymmetric growth rate $- (k_r/2) \langle \tilde{\mathrm{C}} \hat{\Gamma}_1 \rangle$ agrees with the GAM properties observed in Fig. \ref{fig:xpoint_influence} a).
Note that the above calculation also yields a frequency shift due to the asymmetric transport $\delta \omega_{GAM} = k_r^2 \langle \hat{\mathrm{C}} \hat{\Gamma}_2 \rangle$ which is positive because both, $\hat{\mathrm{C}}(\theta)$ and $\hat{\Gamma}_2(\theta)$, are negative (positive) above (below) the outboard midplane \cite{itoh_hallatschek_gam_excitation}.

%
\paragraph{Summary and Conclusions.---}
Compared to linear predictions, numerical ITG turbulence studies display strongly enhanced GAM group velocities in case of non-adiabatic electron response with a dependence on the parallel resistivity.
By analyzing the nonlinear frequency modification we could identify the mechanism responsible for this nonlinear frequency shift. 
The interaction with the turbulence can raise the group velocity of the GAM from the order of the curvature drift velocity ($\sim 0.1$ km/s, $T\sim 100$ eV, $B\sim 1$ T, $R\sim 1$ m) in the linear case \cite{hager_gam,hager_gam_asym} to the order of the diamagnetic drift velocity ($\sim 1$ km/s), which is the typical scale of turbulent motions.
A global nonlinear eigenmode
can be wide enough to form frequency plateaus as observed in ASDEX Upgrade \cite{conway_gamfreq}.
The group velocity determines the flux of the energy stored in the GAM. Therefore, fast propagation might also play a role in considerations about the efficiency of externally driven GAMs as transport barriers.

Up-down asymmetry of the magnetic configuration results in an additional GAM growth rate
causing a preference for one particular sign of the phase velocity. In \textsc{nlet} runs using single-null geometry with the ion curvature drift directed upwards and upper (lower) X-point this interaction favors the excitation of GAMs with negative (positive) group and phase velocities.
Taking into account a radial frequency gradient, the phase velocity preference results in a pulsed activity of turbulence and GAMs for upper X-point.
In contrast, for lower X-point pulsing is subdominant and the turbulence saturates into a quasi-steady state with the mean turbulence intensity being significantly lower than for upper X-point.
The pulsed GAM activity reported here may be related to the pulsed GAM activity observed recently in ASDEX Upgrade \cite{conway_iphase2}. It might also be involved in the explanation of the quiet periods in NSTX \cite{zweben_quiet_periods}. Since the ion curvature drift was directed towards the X-point in all of the discharges analyzed in \cite{zweben_quiet_periods}, it would be interesting to investigate the changes related to an inversion of the curvature drift in the experiment.

\begin{acknowledgments}
 We thank G. Conway and S. Zweben for fruitful discussions.
\end{acknowledgments}


%


\end{document}